\documentstyle[multicol,pre,aps]{revtex}

\newcommand{\r}{\bbox{r}}
\newcommand{\obar}{\displaystyle\overline}
\newcommand{\wtilde}{\displaystyle\widetilde}
\newcommand{\m}{\raisebox{-1mm}{\scriptsize {\textit M}}}

\begin{document}

\draft
\title{ Field Theoretical Representation of\\the Hohenberg-Kohn Free Energy for Fluids }
\author{H. Frusawa\thanks{Electric address: {\ttfamily
furu@exp.t.u-tokyo.ac.jp}} and R. Hayakawa}
\address{Department of Applied Physics, University of Tokyo, Bunkyo-ku, Tokyo 113-8656, Japan}
\date{\today}
\maketitle

\begin{abstract}
To go beyond Gaussian approximation to the Hohenberg-Kohn free energy playing the key role in the density functional theory (DFT), the density functional \textit{integral} representation would be relevant, because field theoretical approach to perturbative calculations becomes available.
Then the present letter first derives the associated Hamiltonian of density functional, explicitly including logarithmic entropy term, from the grand partition function expressed by configurational integrals.
Moreover, two things are done so that the efficiency of the obtained form may be revealed: to demonstrate that this representation facilitates the field theoretical treatment of the perturbative calculation, and further to compare our perturbative formulation with that of the DFT.
\end{abstract}



The Hohenberg-Kohn (HK) free energy---in terms of the density functional theory (DFT) \cite{DFT}---is the natural extension of the Helmholtz one. At the beginning, we see what is meant by this via the primary definition of the HK free energy \cite{DFT}.

Let us start with a grand canonical system which has a volume of $V$ and is surrounded by a reservoir of a chemical potential $\mu$ in the unit of the thermal energy $k_BT$ \cite{KT}.
We consider one-component classical particles, to keep notations as possible as simple, and define the grand potential $\Omega$ in the form:
  \begin{eqnarray}
       e^{-\Omega} = \sum_{N=0}^{\infty} \frac{1}{N!}\,\displaystyle\int
                          \prod_{i=1}^N\,d\r_i
                          \ \exp\;[-\{\sum_{i,j} U(\r_i, \r_j)
                          + \sum_i J(\r_i) - \mu N\}\,],
   \label {eq:omdef}
  \end{eqnarray}
with $\r$ being the position vectors of particles, $J$ the external potential, and $U (\r_i, \r_j)$ the two-body interaction potential for particles $i$ and $j$.
In the thermodynamic limit, the Helmholtz free energy $F$ of a canonical system with $\obar{N}$ particles can be obtained from $\Omega$ by a Legendre transform that $F=\Omega+\mu \obar{N}$, where $\obar{N}$ is the averaged total number given by the relation $\obar{N}=-\partial\Omega/\partial\mu$ in the grand canonical formalism.
Similarly, the HK free energy $F_{HK}$ is defined by replacing $\obar{N}$ in this mapping with the averaged density field $\varphi(\r)$ given in terms of the shifted external field $\wtilde{J}(\r)=J(\r)-\mu$ as
  \begin{equation}
      \varphi = \langle \hat{\rho} \rangle
              = \frac{\delta \Omega}{\delta \wtilde{J}},
   \label{eq:aveddef}
  \end{equation}
where $\hat{\rho} \equiv \sum_{i=1}^N \delta(\r - \r_i )$
is the density {\itshape operator} and $\langle \cdots \rangle$ represents the ensemble average with the weight in Eq. (\ref{eq:omdef}):
the Legendre transform of $\Omega$ with use of $\varphi$ and $\wtilde{J}$ yields
  \begin{equation}
      F_{HK}(\varphi) = \Omega(\tilde{J}) - \varphi\cdot\wtilde{J},
   \label{eq:hkdef}
  \end{equation}
where $\varphi\cdot\wtilde{J}\equiv\int d\r \varphi\wtilde{J}$. Since the relation (\ref{eq:hkdef}) is reciprocal, $F_{HK}$ exactly satisfies the equation that $\delta F_{HK}\,/\,\delta \varphi = -\wtilde{J}$.

In spite of the above generality and tractability of the HK free energy $F_{HK}$, the introduction of $F_{HK}$ has been mostly in the context of the DFT.
Indeed the DFT is one of the most powerful tools for investigating not only spatially inhomogeneous states for simple liquids but also a variety of more complex fluids (e.g. liquid crystals and polymers) \cite{DFT}.
However, it is also to be noted that there are cases where the largely fluctuating systems such as fluids near critical points are beyond its scope. This is seen from the following Ramakrishnan-Youssouff form \cite{RY}:
 \begin{eqnarray}
     \quad\Delta\,F_{HK}^A=\int{d\r}\;
                      \varphi^A(\r)\,\ln \frac{\varphi^A(\r)}{\obar{\rho}_{\m}^A}-\frac{1}{2}\int{d\r}{d\r'}
     \;C^{(2)}(\r-\r';\obar{\rho}_{\m}^A)%
     \Delta\varphi^A(\r)\Delta\varphi^A(\r'),
   \label{eq:hkpertDFT}
  \end{eqnarray}
where $\Delta\,F_{HK}^A$ is the excess free energy around an arbitrary uniform density $\obar{\rho}_{\m}^A$ in an $A$-domain (e.g., liquid region in liquid-vapor coexisting state), $C^{(2)}(\r-\r';\obar{\rho}_{\m}^A)$ is the second-order direct correlation function, and $\Delta\varphi^A=\varphi-\obar{\rho}_{\m}^A$ is the density difference between $\obar{\rho}_{\m}^A$ and $\varphi$ obtained from the self-consistent equation:
  \begin {equation}
    \varphi(\r)=\exp\>[\,C^{(1)}(\r;\varphi)-\wtilde{J}(\r)\,].
   \label{eq:faiDFT}
  \end{equation}
The merit of this form (\ref{eq:hkpertDFT}) is that short scale correlation---particularly crucial for fluids---may be taken into elaborate account via the input of the direct correlation function; this insert is much of benefit due to the extensive study on the Orstein-Zernike integral equation \cite{liquid}.
For all that, the limitation is also to be realized that the direct correlation functions as input effectively consider only quadratic fluctuations in case these are the solutions of the integral equation within the mean spherical approximation including the equivalence, i.e., Percus-Yevick one to hard-core potentials \cite{RPA}.

A systematic way of going beyond the Gaussian approximaion to the HK free energy is to start with the density functional \textit{integral} representation so that field theoretical approach to the perturbative calculations may be available.
Then we purpose, first of all, to transform straightforwardly the primary definition of $F_{HK}$ expressed by configurational integrals to the functional integral representation: (i) to derive from Eq. (\ref{eq:omdef}) with the relation (\ref{eq:hkdef}) the following expression that
\begin{eqnarray}
    e^{-\,F_{HK}}
       &=& \int{D}\rho\ \exp\>[-\{\,H_{sad}(\rho)-\varphi\cdot\wtilde{J}\}\,]
        \label {eq:hkfuncrep}\\
    H_{sad}(\rho)&=&\int{d\r}\,{d\r'}\;
        \frac{1}{2}\,\rho(\r)\,U(\r,\r')\,\rho(\r')+\rho(\r)\wtilde{J}
        +\rho(\r)\ln{\rho(\r)}-\rho({\r}),
        \label{eq:hkHamilrep}
\end{eqnarray}
where $\rho$ denotes an instantaneous quantity (not {\itshape operator}) of density.

The justification of the Hamiltonian (\ref{eq:hkHamilrep}) is the main result of this letter;
one has had few grounds so far for the form (\ref{eq:hkHamilrep}), except for the primitive discussion that division of ideal gas systems into {\itshape cells} produces the entropy term, i.e., the last two terms on the right hand side of Eq. (\ref{eq:hkHamilrep}) \cite{ENT,Zhou}, though the above expression (\ref{eq:hkHamilrep}) comprising the familiar free energy functional is trivial intuitively and hence the corresponding form for the Helmholtz free energy has been often used {\itshape a priori} \cite{Zhou,FIOLD-I,FIOLD-P}.

\begin{flushleft}
(i)
\end{flushleft}

As usual, let us insert into the position vectors' form (\ref {eq:omdef}) of the grand partition function, the identity with use of the auxiliary field $\psi(\r)$: $\int{d}\rho\,{d}\psi\;\exp\>[\,i\psi(\rho-\hat{\rho})\,]=1$.
Then Eq. (\ref {eq:omdef}) reads
  \begin{eqnarray}
    e^{-\Omega}
       &=&\int{D}\rho\,{D}\psi\;
           \exp\>[\,-H\,(\rho,\psi)\,]
        \label {eq:omfuncrep}\\
    H(\rho,\psi)&=&\int{d}\r\,{d}\r'\;
        \frac{1}{2}\,\rho\,U\,\rho'
        +\rho\,J-i\rho\,\psi-\exp\,(-i\psi+\mu),
        \label{eq:omHamilrep}
  \end{eqnarray}
with setting $\rho(\r)=\rho$ and $\rho(\r')=\rho'$.
Previous treatments of Eqs. (\ref{eq:omfuncrep}, \ref{eq:omHamilrep}) have conventionally proceeded to perform Gaussian integration over $\rho$ for fixed $\psi$---the Hubbard-Stratonovich transformation, what is called \cite{HS,CANO}.
The reduction is exact indeed, and hence there appears to be no choice but to do so.
As a matter of fact, however, Gaussian {\itshape approximation}---this also considering quadratic contribution---would be applicable to $\psi$ for given $\rho$: we have the alternatives in mapping Eqs. (\ref{eq:omfuncrep},\ref{eq:omHamilrep}) to more tractable forms.

Then, taking the latter approach, Eqs. (\ref{eq:omfuncrep},\ref{eq:omHamilrep}) are reduced to
  \begin{eqnarray}
    e^{-\Omega}
       = \int{D}\rho\;
           \exp\>[\,-H_{sad}\,]
        \label {eq:omfuncrep2}
  \end{eqnarray}
and $H_{sad}(\rho)$ given by Eq. (\ref{eq:hkHamilrep}); $H_{sad}$ is equal to the Hamiltonian (\ref{eq:omHamilrep}) along the saddle point path for $\psi$, i.e., $H(\rho,\psi_{sad})$ with $\psi_{sad}$ satisfying the saddle point equation that $\delta\,H(\rho,\psi)/\delta\,\psi |_{\psi=\psi_{sad}} =0$.
To be noted in Eq. (\ref{eq:omfuncrep2}), the excess grand potential $\Delta\Omega=-\ln\,[\int\,D\psi\;e^{-\rho(\delta\psi)^2/2}]$ arising from quadratic fluctuations of $\delta\psi=\psi-\psi_{sad}$ is absent. This is due to the following evaluation:
Gaussian integration over $\psi$---carried out by discretized fields, $\rho_l$ and $\psi_l$---yields the apparently nontrivial term of the Lee-Yang type \cite{LY} such that
  \begin{equation}
    \Delta\Omega= \lim_{a^3\to 0} \frac{1}{a^3}
                         \int{d\r}\;\frac{1}{2}\ln\,(\rho_l\,a^3), 
   \label{eq:leeyang}
  \end{equation}
in the continuum limit (or the vanishing limit of the lattice constant $a$ defined as $a^3=(\rho_0)^{-1}\equiv{V}/\obar{N}$), where the other trivial term has been formally absorbed into the integral measure $D\rho$ following the standard procedure \cite{LY}.
The excess potential $\Delta\Omega$ given by Eq. (\ref{eq:leeyang}), however, converges to $0.5(N-\obar{N})$, the half of the difference between the actual total number $N$ and the most probable value $\obar{N}$, as found from the expansion that $\ln(\rho_l\,a^3)=(\rho_l-\rho_0)\,a^3+O[\{(\rho_l-\rho_0)\,a^3\}^2]$. Thus $\Delta\Omega$ is negligible in the thermodynamic limit.

In the next step, we equate the functional derivative $\delta \Omega/\delta \wtilde{J}$ using the representation (\ref{eq:omfuncrep2}) with the averaged density $\varphi$ obtained from both the relation (\ref{eq:aveddef}) and the position vectors' expression (\ref{eq:omdef}) of $\Omega$: we put $\varphi\equiv\langle \hat{\rho} \rangle=\langle \rho \rangle_c$ with $\langle \cdots \rangle_c$ denoting the average under the weight in Eq. (\ref{eq:omfuncrep2}).
Then the Legendre transform (\ref{eq:hkdef}) leads to the expressions (\ref{eq:hkfuncrep}, \ref{eq:hkHamilrep}), in question, of the HK free energy $F_{HK}$;
the principal purpose of this letter has been accomplished.

What we have to do, in addition, would be to show virtues of the now justified representations (\ref{eq:hkfuncrep}, \ref{eq:hkHamilrep}).
Then two things are done in the remainder: (ii) to demonstrate that this form facilitates the field theoretical treatment of the perturbative calculation, and (iii) to compare our perturbative formulation with that of the DFT.

\begin{flushleft}
(ii)
\end{flushleft}

Let us return to the grand potential $\Omega$ given by Eq. (\ref{eq:omfuncrep2}).
The saddle point equation that $\delta{H}_{sad}(\rho)/\delta\,\rho |_{\rho=\rho_{\m},\,J=0} =0$ in the absence of external potential, $J(\r)\equiv{0}$, produces the mean-field density:
  \begin{equation}
     \rho_{\m}(\r)=
     \exp\,\left\{\,-\int{d\r'}\,U(\r,\r')\,\rho_{\m}(\r')+\mu\,\right\}.
    \label{eq:meand}
  \end{equation}
Expanding around $\rho_{\m}$ the logarithmic term in the Hamiltonian difference $\Delta{H}=H_{sad}(\rho)-H_{sad}(\rho_{\m})$, we obtain
  \begin{eqnarray}
    e^{-\Omega}&=&e^{-H_{sad}(\rho_{\m})}
    \int D\wtilde{\rho}\;\exp \{\,-(\Delta\,H+\wtilde{\rho}\cdot J)\,\}
     \label{eq:pertom}\\
    \Delta{H}(\wtilde{\rho})&=&\int{d\r}\,{d\r'}\;
        \frac{1}{2}\>\wtilde{\rho}\,(\,U+\frac{\delta[\r-\r']}{\rho_{\m}}\,)\,\wtilde{\rho}'
        +E(\wtilde{\rho}),
       \label{eq:pertdh}
  \end{eqnarray}
where $\wtilde{\rho}=\rho-\rho_{\m}$, the fluctuation of the total number is neglected (i.e., $\int\,d\r \wtilde{\rho}\approx 0$), and $E(\wtilde{\rho})$ denotes the terms higher than quadratic ones due to the logarithmic expansion. 

The Hamiltonian $\Delta{H}$ given by Eq. (\ref{eq:pertdh}) has some characteristics other than standard field theoretical formulations \cite{LY}:
One is that the free part of $\Delta{H}$ includes the positionally dependent coefficient $1/\rho_{\m}(\r)$, indicating that our formalism is applicable to investigating density fluctuations in structured fluids where the mean-field density itself is not uniform but spatially oscillates.
Next, $\Delta{H}$ reveals that higher terms $E(\wtilde{\rho})$ in one-component systems arise from entropy allowing thermally activated hopping processes and not from interactions \cite{KW}.
Finally it is noted that $\wtilde{J}$ is reduced to $J$.
This is why the external potential $J$ is not included in determining the mean-field density from the saddle point equation; if we take as a reference density the mean-field one in the presence of $J$, $\Delta{H}$ has no external source to create a generating functional.

Of particular interest is the first property; this, for example, makes it possible to generallize Debye-H\"uckel equation as Fisher et al. propose recently \cite{GDH,prep}.
The present letter, however, restricts itself to the simplified systems which have only the short-ranged potentials and consist of macroscopic subsystems (or domains) inside which the mean-field density $\rho_{\m}(\r)$ take constant values except the boundaries.
A typical situation is the coexisting phase of two subsystems (such as liquid and gas).

Consequently, neglect of the interfacial energy reduce $\Delta{H}$ to the sum of the contributions from the domains.
With the volume and the mean field density in an $s$-domain ($s=A,B,\cdots$) denoted by $V_s$  and $\obar{\rho}_{\m}^{\,s}$, respectively (where the overbar is for emphasizing the constancy), the functional integral in Eq. (\ref{eq:pertom}) reads in terms of the Fourier transformed density, $\rho^s(\bbox{k})=(1/V_s)\int{d}\r\exp(i\bbox{k}\cdot\r)\rho^s(\r)$:
  \begin{eqnarray}
    \int D\wtilde{\rho}\;\exp\,&&(-\Delta{H})\nonumber\\
    &&=\prod_{s=A,\cdots}\;C\,\exp\,\left.\left(\,\frac{1}{2}
    \frac{\delta}{\delta\rho^s}\Delta_F^s
         \frac{\delta}{\delta\rho^s}\,\right)\,
        \exp\,\left[\,-\frac{V_s}{(2\pi)^3}\int{d}\bbox{k}\,
        \{E(\rho^s)+\rho^s\,J\,\}\right]
         \right|_{\rho^s=0},
    \label{eq:propa}
  \end{eqnarray}
where further shifted density is abbreviated to $\rho^s$, $C$ is a constant independent of $J$, and the propagator $\Delta_F^s$ is given in the usual form $(\Delta_F^s)^{-1}=u_0+(1/\obar{\rho}_{\m}^{\;s})+u_2\bbox{k}^2$ due to the conventional expansion of the short-ranged potential, $U=u_0+u_2\bbox{k}^2$, in the $\bbox{k}$--space.

The expression (\ref{eq:propa}) implies that Feynman graphs are now available. Thus the HK free energy $F_{HK}$ gets into spotlight, because $F_{HK}$ given by Eqs. (\ref{eq:hkfuncrep},\ref{eq:hkHamilrep}) includes the generating functional of the {\itshape n}-point vertex functionals $\Gamma^{(n)}$ consisting only of one-particle irreducible diagrams as well as the standard field theory \cite{LY}:
putting that $\Delta\varphi^{\,s}=\varphi-\obar{\rho}_{\m}^{\,s}$ in an $s$-domain ($s=A,B,\cdots$) as in Eq. (\ref{eq:hkpertDFT}) and ignoring the number fluctuation (i.e., $\int\,d\r\,\Delta\varphi^{\,s} \approx 0$) as before, $F_{HK}$ reads
  \begin{eqnarray}
    F_{HK}&=&\sum_{s=A,\cdots}F_{MF}^{\,s}+\Delta{F}_{HK}^{\,s}
      \label{eq:perthk}\\
    {F}_{MF}^{\,s}&=&\int{d\r}\,{d\r'}\;
        \frac{1}{2}\,\obar{\rho}_{\m}^{\,s}\,U\,\obar{\rho}_{\m}^{\;'s}
        +\obar{\rho}_{\m}^{\,s}\ln\obar{\rho}_{\m}^{\,s}-\obar{\rho}_{\m}^{\,s}
      \label{eq:pertmf}\\
    \Delta{F}_{HK}^{\,s}&=&-\sum_{n\geq2}^{\infty}\frac{1}{n!}
    \int{d\r_1}\cdots{d\r_n}
    \;\Gamma^{(n)}(\r_1,\cdots,\r_n;\,\obar{\rho}_{\m}^{\,s})\,\Delta\varphi^{\,s}(\r_1),\cdots,\Delta\varphi^{\,s}(\r_n).
    \label{eq:gamma}
  \end{eqnarray}
It is to be noted in the above representations that $F_{HK}$ correctly includes the mean-field Helmholtz free energy $F_{MF}$ in the absence of external potential $J$, in contrast to previous density functional integral formulations \cite{CANO} which take as a reference the free energy for the density smeared over the entire system. As a result, the present formalism may focus on appropriate density deviation from not a smeared value but a mean-field one $\obar{\rho}_{\m}^{\,s}$, and therefore makes the perturbative approximation more precise than previous approaches.

\begin{flushleft}
(iii)
\end{flushleft}

Even at the starting point, there exists difference between the DFT and our formalism: the former separates the entropy term, $\int{d\r}\,\varphi\ln\varphi-\varphi$, from ${F}_{HK}$ beforehand, whereas the latter does the mean-field free energy $H_{sad}(\rho_{\m})$ (see Eq. (\ref{eq:pertom})).
Such distinct types of theories, though, have correspondence in some cases as will be seen below.

Most successful is the absence of interactions between particles, i.e., $U=0$, where the field theoretical representation is identified with the DFT formalism. To see this, we consider the grand potential $\Omega$.
In the DFT, $\Omega$ is reduced to $\Omega=\int{d\r}\,\varphi_{\mathrm{ex}}\ln\varphi_{\mathrm{ex}} + \varphi_{\mathrm{ex}}\wtilde{J}$ with the averaged density $\varphi_{\mathrm{ex}}=\exp(-\wtilde{J})$ obtained from the relation (\ref{eq:faiDFT}) when $C^{(1)}=0$.
While the saddle point equation, $\delta{H}_{sad}/\delta\rho=0$, explicitly including the external potential $J$ also produces the density equal to $\varphi_{\mathrm{ex}}$, and therefore the deviation of $\Omega$ in the functional integral formalism from that for the DFT is given in the first evaluation as $-\ln\,[\int{D\,\delta\rho}\,\exp\,(\,-{(\delta\rho)^2}/{2\varphi_{\mathrm{ex}}}\,)\,]$, being the same kind as the excess grand potential $\Delta\Omega$.
Then repeating the similar discussion to that after Eq. (\ref{eq:leeyang}), the difference may be ignored and thus the equivalence in the case of $U=0$ is assured.

For $U\neq 0$, on the other hand, let us compare both representations of the excess HK free energy $\Delta{F}_{HK}^{A}$ in an $A$-domain up to quadratic terms for $\Delta\varphi^{A}$.
For the DFT formulation (\ref{eq:hkpertDFT}), the expansion of the logarithmic term yields
  \begin{eqnarray}
    \Delta\,F_{HK}^{A}=\int{d\r}\,{d\r'}\;
     \frac{1}{2}\left(\,-C^{(2)}+\frac{\delta[\r-\r']}{\obar{\rho}_{\m}^{A}}\,\right)
        \,\Delta\varphi^{A}(\r)\Delta\varphi^{A}(\r').
   \label{eq:mhkdf2}
  \end{eqnarray}
In applying the mean spherical approximation (MSA) to the calculation of the direct correlation function for hard-core(-Yukawa) or square well fluids \cite{liquid}, we have only to put $-C^{(2)}=U$ in Eq. (\ref{eq:mhkdf2}), because the other condition that the pair distribution function is to be set to zero inside hard spheres with the diameter $d$ is formally satisfied due to the potential of $U=\infty$ for $\r-\r'\leq{d}$.
On the other hand, the functional integral representation (\ref{eq:gamma}) under the tree approximation reduces to $\Delta\,F_{HK}^{A}=\Delta{H}(\Delta\varphi^A)$  \cite{LY}, and therefore takes the same form up to quadratic terms as the DFT expression (\ref{eq:mhkdf2}) with use of the MSA for the above mentioned fluids (see also Eq. (\ref{eq:pertdh})).

The conformity for $U\neq 0$ contrarily results in highlighting some merits of the field theoretical forms (\ref{eq:perthk}) to (\ref{eq:gamma}), which we describe in conclusion.
One virtue other than the DFT using the MSA is that the density functional {\itshape integral} representation may still take fluctuations into more elaborate consideration systematically, by including loop graphs in Eq. (\ref{eq:gamma}).
Moreover, we would like to stress that a reference density---this being merely an arbitrary value in the DFT (see the statement just after Eq. (\ref{eq:hkpertDFT}))---is obtained from the relation (\ref{eq:meand}) self-consistently in our formalism.

\bigskip
We acknowledge the financial support from the Ministry of Education, Science, Culture, and Sports of Japan under Grant No. 10450032.



\begin{references}
\bibitem{KT}
Hereafter, all of {\itshape potentials} ($\mu,\,\Omega,\,J,\,U,\,F$, and $H$) are in the $k_BT$ unit.
\bibitem{KW}
The two-body interaction potential $U$ in the case of one-component systems is not averaged but bare one. Therefore the pairwise additivity does not mean the Kirkwood superposition and would not be so crude approximation.
\end{references}
\end{document}